\documentclass[pra,showpacs,preprintnumbers,amsmath,amssymb,tightenlines,twocolumn]{revtex4-1}

\usepackage{graphicx}
\usepackage{dcolumn}
\usepackage{bm}
\usepackage{epsfig}
\usepackage{amsmath}

\newcommand{\bra}[1]{\langle\,{#1}\, |}
\newcommand{\ket}[1]{|\,{#1}\,\rangle}

\newcommand{\sub}[2]{{#1}_{\mbox{\!\! \scriptsize #2}}}
\newcommand{\supp}[2]{{#1}^{\mbox{\!\! \scriptsize #2}}}

\def\beq{\begin{equation}}
	\def\eeq{\end{equation}}

\newcommand{\fref}[1]{Fig.~\ref{#1}}

\newcommand{\eref}[1]{Eq.~(\ref{#1})}

\newcommand{\cref}[1]{chapter~\ref{#1}}

\newcommand{\Cref}[1]{Chapter~\ref{#1}}

\newcommand{\aref}[1]{appendix~\ref{#1}}

\usepackage{ulem}  
\usepackage{color}

\begin{document}
	\title{Influence of disordered and anisotropic interactions on relaxation dynamics and propagation of correlations in tweezer arrays of Rydberg dipoles}
	\author{K.~Mukherjee}
	\affiliation{Homer L. Dodge Department of Physics and Astronomy, The University of Oklahoma, Norman, Oklahoma 73019, USA}
	\affiliation{Center for Quantum Research and Technology, The University of Oklahoma, Norman, Oklahoma 73019, USA}
	\author{G.~W.~Biedermann}
	\affiliation{Homer L. Dodge Department of Physics and Astronomy, The University of Oklahoma, Norman, Oklahoma 73019, USA}
	\affiliation{Center for Quantum Research and Technology, The University of Oklahoma, Norman, Oklahoma 73019, USA}
	\author{R.~J.~Lewis-Swan}
	\affiliation{Homer L. Dodge Department of Physics and Astronomy, The University of Oklahoma, Norman, Oklahoma 73019, USA}
	\affiliation{Center for Quantum Research and Technology, The University of Oklahoma, Norman, Oklahoma 73019, USA}
	\email{lewisswan@ou.edu}

\begin{abstract}
        We theoretically investigate the out-of-equilibrium dynamics of irregular one- and two-dimensional arrays of Rydberg dipoles featuring spatially anisotropic interactions. Starting from a collectively polarized initial state, we map out the dynamical phase diagram and 
        identify a crossover between regimes of regular and anomalously slow relaxation of the initial collective order, that strongly depends on both the degree of interaction disorder and anisotropy. In addition, we find the regime of slow relaxation is characterized by a sub-ballistic propagation of correlations that remained confined to short distances even at long times. To explain our findings we develop an analytic model based on decoupled clusters of interacting dipoles that goes beyond prior theoretical works and enables us to identify multiple relaxation timescales. Our findings can be relevant for a wide variety of quantum science platforms naturally featuring disordered dipolar interactions, including polar molecules, frozen Rydberg gases and NV centers.
\end{abstract}
	
\maketitle
 
\section{Introduction} Highly excited Rydberg atoms in programmable tweezer arrays are attracting a significant amount of interest for quantum science applications \cite{adams2019rydberg}, including analog quantum simulation \cite{weimer2010rydberg,bernien2017probing,levine2018high,browaeys2020many,kaufman2012cooling,madjarov2020high,scholl2021quantum,morgado2021quantum,kim2018detailed}, digital quantum computing \cite{cohen2021quantum,wurtz2023aquila, Mitra2023} and entanglement-enhanced metrology \cite{eckner2023realizing,madjarov2020high,morgado2021quantum,finkelstein2024universal,cao2024multi,Brif2020}.
In the context of quantum simulation, arrays of Rydberg atoms can be used to simulate a variety of tunable models due to their strong dipolar interactions and long coherence times. For example, encoding qubits in a pair of Rydberg states or a combination of Rydberg and ground
states
enables the realization of tunable XY \cite{bornet2023scalable,orioli2018relaxation}, XXZ \cite{signoles2021glassy,scholl2022microwave} and Ising \cite{schauss2018quantum,labuhn2016tunable} models of quantum magnetism as well as complex quantum systems such as light-harvesting complexes \cite{David:Rydagg,mukherjee2020two}.

The maturing combination of Rydberg atoms with programmable tweezer array architectures has distinguished them from, e.g., analogous simulation of dipolar spin models using qubits encoded in rotational states of polar molecules pinned in a deep optical lattice \cite{ospelkaus2006ultracold,ni2018dipolar,yan2013observation,hazzard2014many}. In those cases, the limited preparation efficiency of the molecules results in lattices with a low ($\ll 1$) filling fraction and thus they naturally realize (positionally) disordered quantum spin models \cite{hazzard2014many,hazzard2014quantum,christakis2023probing}. Similar recent work using bulk frozen Rydberg gases \cite{signoles2021glassy,scholl2022microwave,franz2022observation} -- which feature positional disorder due to stochastic excitation of atoms to Rydberg states -- has demonstrated that such disordered quantum spin models can feature slow or \emph{glassy} dynamics, and thus they can provide insight into disordered materials \cite{binder1986spin,phillips1996stretched,gotze1992relaxation}. The glassy relaxation is characterized by a decay of initially imprinted collective order that follows the form of a stretched-exponential \cite{kohlrausch1854theorie}, and is understood to arise from the coherent dephasing of many isolated pairs of interacting spins featuring a range of length and time scales due to positional disorder \cite{schultzen2022glassy,schultzen2022semiclassical}. A drawback in both the case of polar molecules and frozen Rydberg gases is that the positional disorder underpinning the dynamics uncontrollably fluctuates from shot-to-shot of the experiment, which limits the understanding these platforms can directly provide about, e.g., the importance of the statistical distribution of disorder or the role that specific configurations play in the thermalization of the many-body system. 

	\begin{figure}[htb]
		\centering
		\epsfig{file=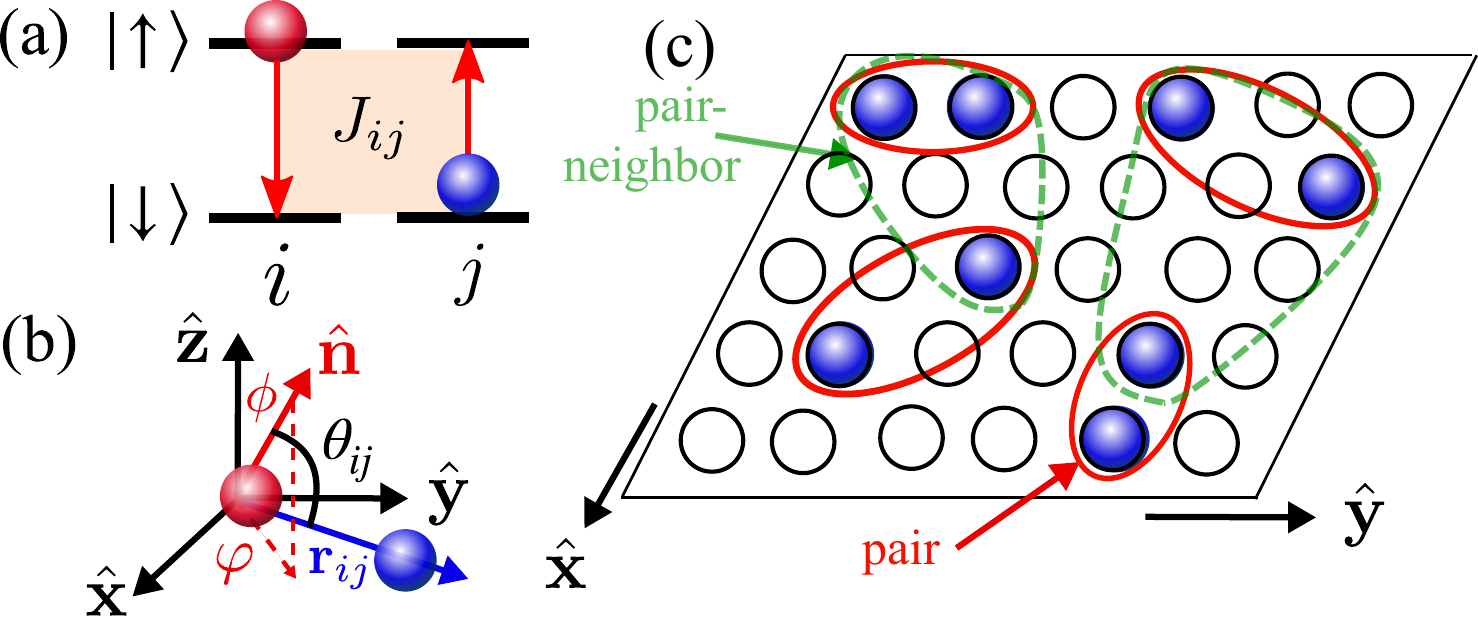,width= 0.99\linewidth}
		\caption{(a) Level-scheme of two Rydberg qubits ($i$ and $j$) with dipolar exchange interactions. The qubit states can be encoded in, e.g., $\ket{nP}$ and $\ket{nD}$ Rydberg states of different parity, where $n$ is the principal quantum number. (b) Sketch of the orientation of a pair with respect to the quantization axis ($\hat{\mathbf{n}}$), which governs the anisotropy in interactions. (c) A 2D array of Rydberg qubits interacting via dipole-dipole interactions. Red circles indicate pair-wise/strongest neighbor coupling and green circles represents pair-neighbor interactions i.e.~pair interacting with additional spin in the array. 
			\label{sketch_setup}}
	\end{figure}

In this context, the ability to prepare tweezer arrays of Rydberg atoms with arbitrary control of spatial positioning \cite{nogrette2014single,barredo2016atom,endres2016atom,barredo2018synthetic}, and thus the capability to deterministically introduce defects, presents an ideal quantum simulation platform to controllably study relaxation dynamics of a disordered spin models. 
As a first step in this direction, we theoretically investigate the dynamical phase diagram of the disordered XY model in one (1D) and two (2D) dimensional arrays and identify the level of positional disorder required for the system to crossover into the glassy regime.
Further, studying the dynamics in 2D arrays provides an ideal setting to controllably interpolate between isotropic and anisotropic dipolar interactions and demonstrate their effect on the long-time dynamics, such as the presence or absence of a prethermal regime.  
Most importantly, we find that the simplistic description of the decay of collective observables in the glassy regime in terms of isolated pairs fails for the dipolar XY model, particularly in the case of 2D arrays. Instead, we develop an extended model involving clusters of three spins that agrees well with our numerical simulations of the full many-body system. 
We also go beyond prior theoretical and experimental investigations by showing that two-body correlations can reveal signatures of the positional disorder, including sub-ballistic propagation of quantum information through the array and suppressed correlations at large length scales even at long times. 
Our findings suggest that higher-order correlations can be an important probe for developing a complete, many-body understanding of slow relaxation in disordered spin systems.



\section{Dipolar XY model in a Rydberg array}\label{sec:Rydberg_XY} To ground our investigation relative to prior experiments with molecules and frozen Rydberg gases, we consider a fictitious lattice of $L^d$ equally spaced sites where $d$ is the dimension of the lattice and $L$ is the number of sites along each dimension. The filling fraction $f = N/L^d$ of the array is controlled by (deterministically) placing $N$ individual tweezers onto lattice sites, each loaded with a single Rydberg atom\cite{nogrette2014single,barredo2016atom,endres2016atom,barredo2018synthetic}. As an example, a 2D lattice with eight Rydberg atoms is illustrated in \fref{sketch_setup}~(c). 

A qubit is encoded in each atom by selectively populating a pair of Rydberg states of different parity, which we will represent 
by $\ket{\downarrow}$ and $\ket{\uparrow}$, as depicted in \fref{sketch_setup}~(a). 
This encoding leads to resonant dipole exchange interactions between pairs of Rydberg qubits, which can be described by an XY Hamiltonian:	
\begin{equation}
    \sub{H}{XY} = \sum_{i>j} \frac{J_{ij}}{4}(\hat{\sigma}^{x}_{i} \hat{\sigma}^{x}_{j} + \hat{\sigma}^{y}_{i} \hat{\sigma}^{y}_{j} ) ,
    \label{Hamil_XY_1}
\end{equation}
where $\hat{\sigma}^a_{k}$ for $a\in\{x,y,z\}$ are Pauli matrices acting on the $k$th atom in the array.
Here, $J_{ij}=-J(1-3\cos^2 \theta_{ij})/r_{ij}^3$ is the dipole-dipole coupling between spins $i$ and $j$ separated by a distance $r_{ij}$, with $J=C_3/a_0^3$ where $C_3$ is the dipole-dipole dispersion coefficient and $a_0$ defines the fictitious lattice constant. 
Note that the absorption of the length scale $a_0$ into the definition of $J$ ensures that the distance $r_{ij}$ is dimensionless (i.e., nearest-neighbour sites are separated by unit distance). 
The anisotropy of the interaction is controlled by $\theta_{ij}$, defined as the angle between the polarization of the dipoles [set by the quantization axis ($\hat{\mathbf{n}}$) determined by a background magnetic field] and the interatomic vector $\mathbf{r}_{ij}$, as shown in \fref{sketch_setup}~(b).

\begin{figure}[tb]
	\centering
	\epsfig{file=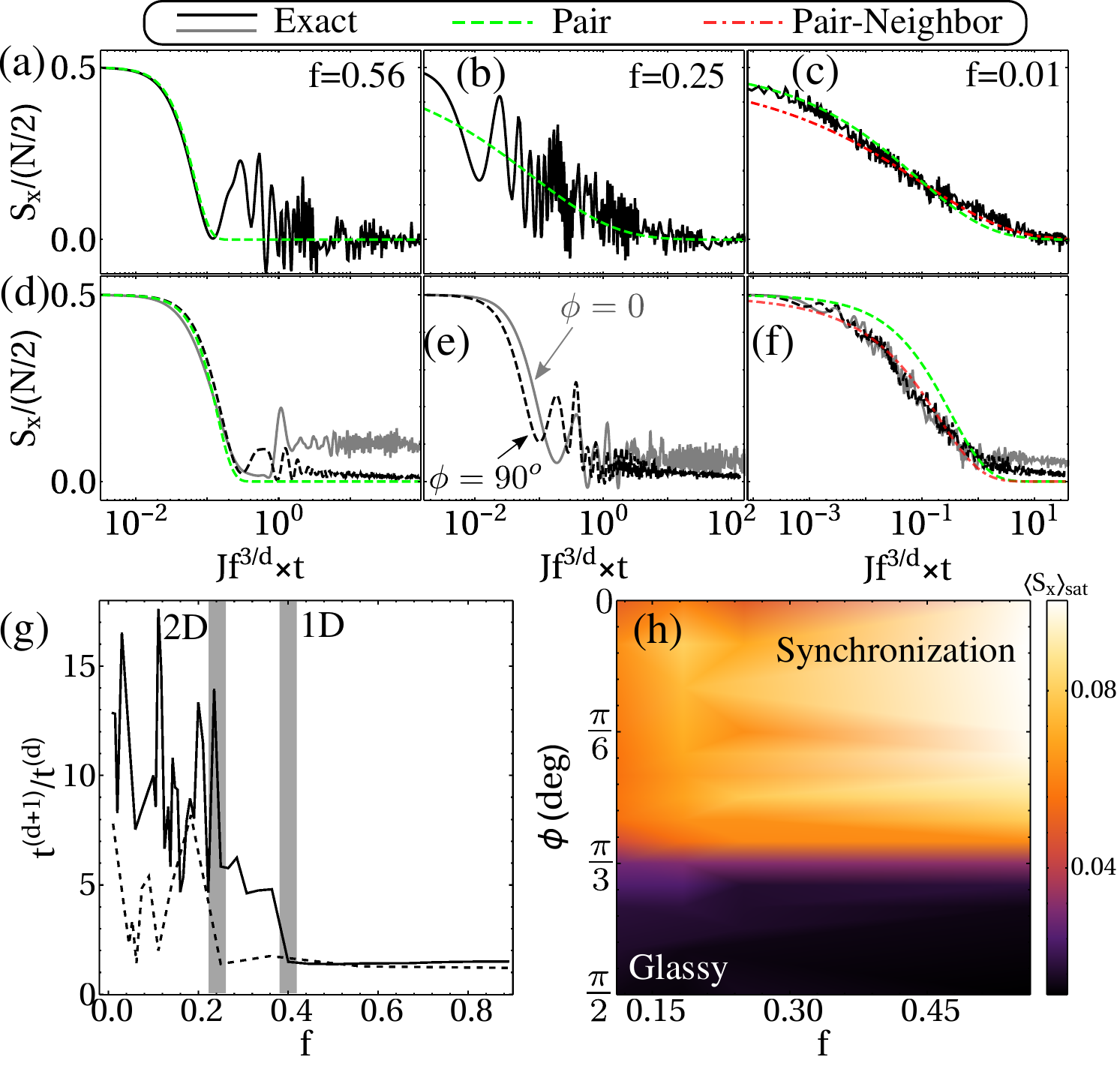,width= 1.05\linewidth}
	\caption{(a)-(f) Time-evolution of ensemble average polarization in 1D array (black-solid) for filling fractions (a) $f=0.56$, (b) $f=0.25$, and (c) $f=0.01$ with $N=6$, and 2D array with isotropic (gray-solid) and anisotropic (black-dashed) dipolar interactions for filling fractions (d) $f=0.56$, (e) $f=0.25$, and (f) $f=0.01$ with $N=9$. The green-dashed line and red-dot-dashed line are stretched exponential functions from the pair model ($\langle \hat{S}_x \rangle_{\mathrm{pair}}$) and pair-neighbor model ($\langle \hat{S}_x \rangle_{pn}$), respectively, in (c) and (f). (g) Dimensionless ratio $t^{(d+1)}/t^{(d)}$ with filling fraction $f$, calculated from relaxation dynamics of 1D (solid) and 2D (dashed) arrays, respectively, see text. (h) Long-time value ($Jf^{\frac{3}2}t\approx 10^2)$ of the polarization $\langle \hat{S}_x \rangle_{\mathrm{sat}}$ as a function of quantization axis polar angle $\phi$ and filling fraction $f$.
		\label{relaxation}}
\end{figure}
	%
	
\section{Dynamical phase diagram with disorder}\label{sec:dynamical_phase_diagram_disorder} We study quench dynamics from an initial product state of all spins collectively polarizing along $\hat{x}$, $\ket{\rightarrow}_x^{\otimes N}=2^{-N/2}(\ket{\uparrow} + \ket{\downarrow})^{\otimes N}$,  and track the relaxation of the collective coherence $\langle \hat{S}_x \rangle=\sum_i^N \langle \hat{\sigma}_i^x\rangle/2$ as a function of filling fraction, where $\langle \hat{\sigma}_i^x\rangle$ represents average over the quantum state of the many-body spin system. The time-evolution of the system is obtained by numerical integration of the Schr{\"o}dinger equation for all $N$ qubits. As our numerical calculations are restricted to relatively small system sizes that preclude self-averaging, we sample between $500$-$1000$ random tweezer configurations and then compute disorder-averaged expectation values. 

In \fref{relaxation}~(a-c), we show representative timetraces of $\langle\hat{S}_x\rangle$ for $f=0.56$, $0.25$ and $0.01$ in a one-dimensional (1D) array. In this case, the dipolar interactions are trivially isotropic [i.e., $\theta_{ij}$ is constant in \eref{Hamil_XY_1}]. For ordered systems with high filling fraction [see panel (a) for $f=0.56$], the collective coherence decays relatively quickly following a Gaussian profile  $\propto \exp[-(\gamma t)^2]$ [green dashed-line] \cite{bornet2023scalable}. For smaller filling fractions [see panels (b-c) with $f=0.25$ and $0.01$], increasingly strong positional disorder leads to qualitatively slower decay. Prior work has described the relaxation in the regime of vanishing $f$ using a so-called pair model where the array can be decomposed into independent pairs of spins, determined by identifying the strongest interacting neighbour for each spin, as shown with red circles in \fref{sketch_setup}~(c), (``weaker'' interactions with additional spins etc are neglected) \cite{schultzen2022glassy}. The dynamics of each pair can be analytically solved, and the resulting collective coherence is obtained by summing over the entire (large) system -- equivalent to averaging over all possible pair configurations -- to yield
a stretched exponential function $\langle \hat{S}_x \rangle_{\mathrm{pair}}=\frac{N}{2}\exp[-(\gamma_p t)^{\beta_p}]$ \cite{kohlrausch1854theorie}. In this expression, for a generic array in $d$ dimensions with power-law interactions defined by $1/r_{ij}^{\alpha}$ we have the characteristic decay rate $\gamma_p \propto Jf^{\alpha/d}$  and stretch exponent $\beta_p = d/\alpha$ \cite{schultzen2022glassy}. Therefore, for a 1D array ($d=1$) with dipolar interactions ($\alpha=3$), the pair model predicts $\beta_p=1/3$. Dynamics with $\beta < 1$ are generically referred to as ``glassy'', indicating anomalously slow relaxation as a result of the disordered couplings between qubits \cite{binder1986spin,phillips1996stretched,gotze1992relaxation}. We comment that an interesting feature of the pair model is that it is somewhat agnostic of the precise details of the spin model, in the sense that the prediction for the stretch exponent is universal for XY, Ising and XXZ spin models \cite{franz2022observation}. 
Comparing the numerical data of the full array to the pair model [solid black and green-dashed lines in panel (c), respectively] we find reasonable agreement in the stretch exponent, although we consider $\gamma$ as a free parameter here for fitting the stretched exponential function to the numerical data, which yields $\gamma = 2\gamma_p$.



We also map out the dynamical phase diagram as a function of filling fraction by quantifying the relaxation timescales with the ratio $t^{(d+1)}/t^{(d)}$, which compares the time taken for $\langle \hat{S}_x \rangle/N$ to reach the fractional values $t^{(d+1)}=0.5/\exp{(d+1)}$ and  $t^{(d)}=0.5/\exp{(d)}$, where $d$ is the dimension of the array. The black solid line shows this ratio for $1$D in \fref{relaxation}~(g).  
In the ordered regime, we expect Gaussian decay of $\langle \hat{S}_x\rangle$ to yield $t^{(d+1)}/t^{(d)}=\sqrt{2}$, which is consistent with that extracted from the numerical data for $f \geq 0.4$. On the other hand, the ratio of timescales clearly detects a dynamical crossover into the disordered regime for $f < 0.4$ via a sudden increase in $t^{(d+1)}/t^{(d)}$, indicating a relative slowdown in relaxation (i.e., a longer tail in $\langle \hat{S}_x \rangle$). From the pair model, we expect $t^{(d+1)}/t^{(d)}=\sqrt[\beta]{d+1/d}$ such that $t^{(d+1)}/t^{(d)} > 2$ for glassy dynamics with $\beta < 1$. The noise in the numerical data is primarily due to the large temporal fluctuations in the decay of $\langle \hat{S}_x \rangle$ due to the finite filling fractions we use, but we clearly observe $t^{(d+1)}/t^{(d)} > 2$ (for clarity, one expects $t^{(d+1)}/t^{(d)} = 8$ for $\beta = 1/3$).

	
For 2D arrays, one can observe richer relaxation dynamics as the relative orientation of the quantization axis [see \fref{sketch_setup}~(b)] can be varied to tune between isotropic ($\phi=0$), achieved by fixing the magnetic field perpendicular to the plane of the 2D array, and anisotropic ($\phi=\pi/2$) dipolar interactions, maintained by aligning the magnetic field along the plane (say $\hat{\mathbf{x}}$ with $\varphi=0$). Here, we consider $\phi$ and $\varphi$ as the polar and azimuthal angles of the quantization axis, respectively as illustrated in \fref{sketch_setup}~(b). We present the relaxation dynamics with isotropic (grey-solid) and anisotropic (black-dashed) interactions in \fref{relaxation}~(d-f). There are two characteristic features exclusive to 2D arrays. Firstly, with isotropic interactions (grey-solid), the system saturates to a residual polarization at long times. This  stems from many-body \emph{synchronization} of the dipoles in 2D that generates a long-lived prethermal state, even in the presence of appreciable positional disorder\cite{kwasigroch2017synchronization}. Conversely, anisotropy in the interactions prevents synchronization and the collective polarization simply relaxes towards zero. We map out these two distinct behaviours in \fref{relaxation}~(h), which shows the saturation value of the polarization, $\langle \hat{S}_x \rangle_{\mathrm{sat}}$,  at long times as a function of the orientation of the quantization axis $(\phi)$ and filling fraction $f$. A clear boundary separates the synchronized regime ($\langle \hat{S}_x \rangle_{\mathrm{sat}} \neq 0$) from complete relaxation ($\langle \hat{S}_x \rangle_{\mathrm{sat}} = 0$) when the quantization axis crosses through $\phi\approx55^o$ for the values of $f$ shown. The presence of the many-body synchronized regime despite appreciable positional disorder is already indicative that a simple description of the dynamics in the disordered regime in terms of isolated pairs is insufficient for $2$D arrays, though the synchronization is known to vanish as $f \to 0$ \cite{kwasigroch2017synchronization}. However, for simplicity we focus on the case of anisotropic interactions in the following discussion of the crossover between ordered and glassy relaxation.


At a high level, the relaxation dynamics for anisotropic interactions in $2$D arrays follows similarly to the prior $1$D case. For large filling fractions [see black dashed line in \fref{relaxation}~(d) for $f = 0.56$] the system exhibits a Gaussian relaxation of the collective polarization. This gives way to a regime featuring slower relaxation as $f$ decreases [see dashed solid lines in \fref{relaxation}~(e-f) for $f=0.25$ and $0.01$, respectively], though a systematic mapping of the dynamical phase diagram using the ratio of timescales $t^{(d+1)}/t^{(d)}$ [see dotted black line in \fref{relaxation}~(g)] shows that the onset of glassy relaxation occurs at a smaller value of $f \approx 0.2$ [shaded region in same panel] relative to $1$D, which we attribute to increased connectivity in $2$D. 

Despite the apparent similarities to $1$D, a closer examination of the relaxation dynamics in the limit of strong positional disorder [see panel (f) for $f = 0.01$ data] reveals a more sophisticated picture. A comparison of the full numerical calculations for $\langle \hat{S}_x \rangle$ with the previously established pair model (green-dashed line) shows significant discrepancies both at short times ($10^{-3}<Jf^{3/2}t<10^{-1}$) and long times ($Jf^{3/2}t>1$). In fact, relaxing the pair model by treating the relaxation rate $\gamma$ and stretch exponent $\beta$ as free parameters and fitting the numerical data, we find $\beta\approx0.43\neq{\beta}_{p}$, where ${\beta}_p=2/3$. 

These deviations emphasize that the simple description of the relaxation dynamics in terms of decoupled pairs of interacting spins is insufficient for the XY model in two dimensions. While it is tempting to attribute this to the dipolar interaction being quasi-long-range in $2$D (as opposed to short-range in $1$D) \cite{kwasigroch2014bose,kwasigroch2017synchronization,defenu2023long}, we find this is not the case. In Appendix~\ref{app:pair} we systematically extract fitted values for $\gamma$ and $\beta$ for strongly disordered arrays in $1$D, $2$D and $3$D as a function of interaction range $\alpha$, and find deviations from the prediction $\beta = d/\alpha$ for all cases in $2$D and $3$D. Instead, we argue that the pair model is fundamentally inappropriate for the XY model, even in the limit $f\to0$. To understand this, we highlight that the pair model was original developed as a solution for glassy dynamics of the Ising model \cite{schultzen2022glassy}. In that case, the Hamiltonian can be formally decomposed into a sum of commuting terms describing interacting pairs, each of which can be independently solved and then recombined with $f \to 0$. Conversely, the XY model is composed of non-commuting terms so the factorization of the array of interacting spins into independent pairwise interactions is strictly never valid. 


Nevertheless, as we show here, we can still gain some analytic insight into the disordered regime by minimally extending the pair description to a \emph{pair-neighbor} model. In this description, we consider a cluster of three spins i.e. the original pair supplemented with an additional neighboring spin, illustrated in \fref{sketch_setup}~(c) with green circles. Solving the pair-neighbor model with some approximations (see Appendix \ref{app:three-spin}) gives an expression for the collective polarization composed of two exponents stretched exponential: $\langle \hat{S}_x \rangle_{pn}=\frac{N}2(\exp[-(\gamma_p t)^{\beta_p}-(\gamma_{pn} t)^{\beta_{pn}}])$, where $\gamma_p$ ($\gamma_{pn}$) and $\beta_p$ ($\beta_{pn}$) are the decay rate and stretch exponents from pair ($p$) and pair-neighbor ($pn$) models, respectively. In general, we find $\beta_p=d/\alpha$ and $\beta_{pn}=d/(2\alpha)$, or $2/3$ and $1/3$ in $2$D with dipolar interactions.  

We find the pair-neighbour model yields a much improved description of the glassy regime in $2$D (red-dot-dashed line in \fref{relaxation}~(f)), particularly across the important time period of $10^{-2} \lesssim Jf^{3/2}t \lesssim 10^0$ where most of the relaxation of the polarization occurs. We attribute discrepancies at shorter timescales to approximations used to obtain a simple analytic form for $\langle \hat{S}_x\rangle_{pn}$ (see Appendix \ref{app:three-spin}), while capturing the dynamics at long times naturally requires capturing weaker interactions with additional spins. For completeness, we also plot the predictions of the pair-neighbour model alongside the results for the case of $f = 0.01$ in $1$D [see red-dot-dashed line in \fref{relaxation}~(c)].
We find the relaxation for $Jf^{3}t \gtrsim 10^{-2}$ is better captured by this extended description as opposed to the original pair model, with the same caveats about the disagreement of our analytic expression at shorter times.  

\section{Propagation of correlations}\label{sec:correlation_propagation} To obtain finer grained information about the relaxation of the many-body quantum state we also examine the dynamics of site-resolved pair correlation functions. Specifically, we compute the lattice-averaged correlation between populations at sites separated by $\mathbf{\delta r}$ \cite{christakis2023probing},
\begin{eqnarray}
		C(\mathbf{\delta r},t) &=& \frac{1}{fL^d}\sum_\mathbf{r} (\langle \hat{n}^{\leftarrow}_{\mathbf{r}} \hat{n}^{\leftarrow}_{\mathbf{r}+\mathbf{\delta r}}\rangle - \langle \hat{n}^{\leftarrow}_{\mathbf{r}} \rangle \langle \hat{n}^{\leftarrow}_{\mathbf{r}+\mathbf{\delta r}}\rangle) . \label{correlation}
\end{eqnarray}
Here, $\langle \cdot \rangle$ implicitly denotes averaging over the quantum state of the many-body spin system as well as ensemble of disorder realizations. The population at lattice position $\mathbf{r}$ is given by $\hat{n}^{\leftarrow}_\mathbf{r}=1/2(\hat{\mathbb{I}} - \hat{\sigma}_x)_\mathbf{r}$ and is computed in a rotated qubit basis, defined by $|\leftarrow \rangle=1/\sqrt{2}(|\downarrow\rangle -|\uparrow \rangle)$. We note that alternative correlations have been studied in $1$D disordered systems \cite{Baldwin_2023_disorderLR}, but we choose Eq.~(\ref{correlation}) as it is well defined for arbitrary dimension arrays.

\begin{figure}[tb]
		\centering
		\epsfig{file=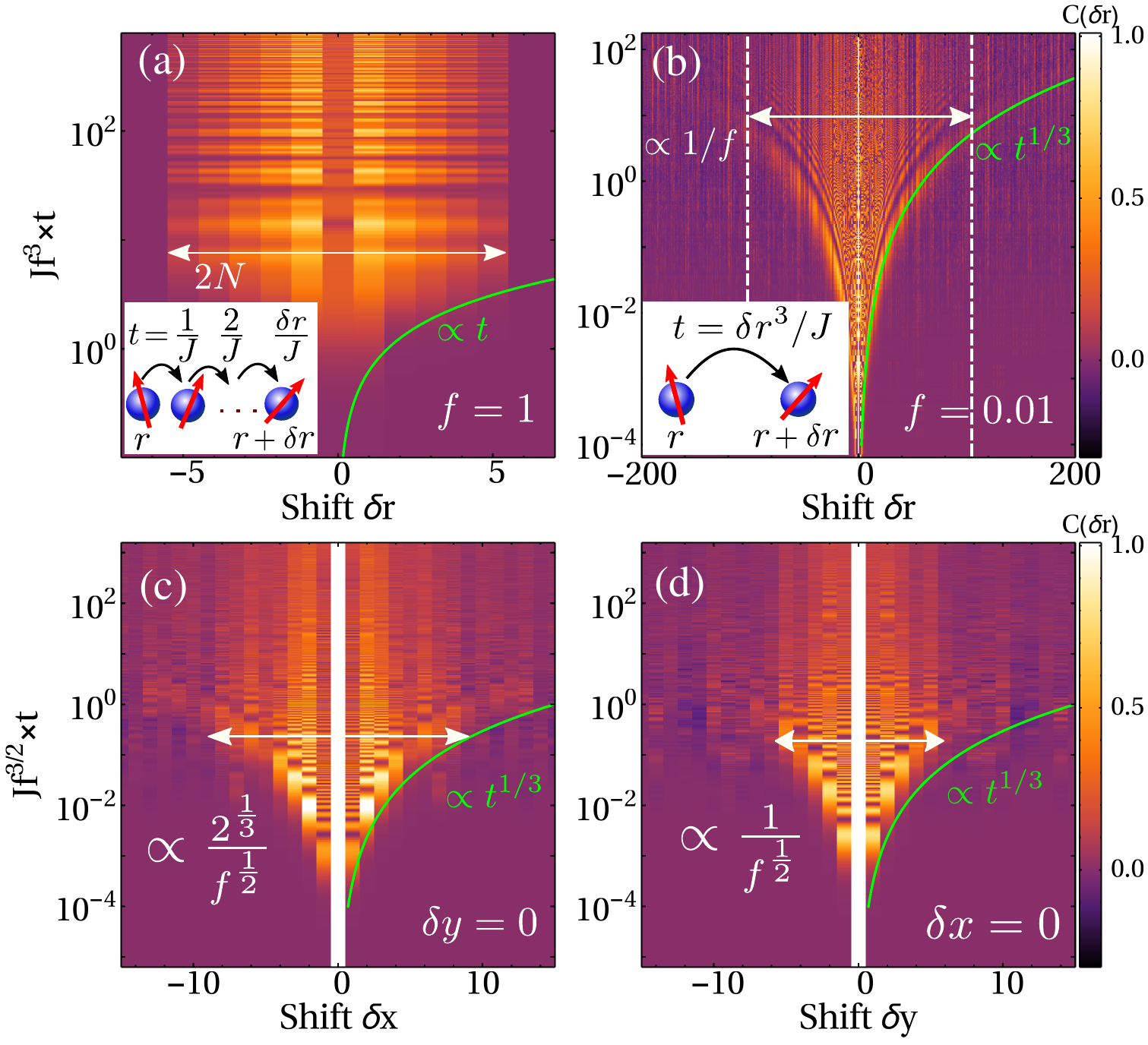,width=1.03 \linewidth}
		\caption{Correlation $C(\mathbf{\delta r},t)$ as function of re-scaled time $Jf^{3/d}t$ and intersite distance. Top row shows numerical data for $1$D arrays with filling fraction (a) $f=1$ ($N=8)$, and (b) $f=0.01$ ($N=6)$. Bottom row is for 2D array with $f=0.01$ $(N=9)$ and orientation of quantization axis ($\phi=\pi/2,\varphi=0$), showing correlations computed between sites separated along the (c) $x$-axis ($\mathbf{\delta r} = \delta x \hat{\mathbf{x}}$) and (d) $y$-axis ($\mathbf{\delta r} = \delta y \hat{\mathbf{y}}$). The different widths along $x$- and $y$-axis are attributed to anisotropic interactions (see main text).
			\label{correlations}}
	\end{figure}

Figure \ref{correlations} shows the evolution of $C(\mathbf{\delta r},t)$ across a one-dimensional array as a function of separation distance $\delta r = \vert \delta \mathbf{r}\vert$ with filling fraction $f=1$ [panel (a)] and $f=0.01$ [panel (b)]. In the former ordered case, correlations propagate ballistically and are confined within a linear ($\propto t$) light cone consistent with well established Lieb-Robinson bounds \cite{lieb1972finite,lucas2021LR,richerme2014non,cheneau2012light}. Specifically, as the dipolar interactions are short-ranged in $1$D, correlations between sites separated by a distance $\delta r$ are primarily mediated through sequential nearest-neighbor interactions between the intervening spins and thus become appreciable at the timescale $t\propto J^{-1}\delta r$. For later comparison, note the correlations eventually (i.e., after $Jt \gtrsim N$) propagate across the entire lattice of $L = N$ sites.

In contrast, strong positional disorder [panel (b)] leads to a slowdown and we observe that correlations propagate sub-ballistically up to timescales $Jf^3 t \lesssim 1$. This is explained by noting that in a dilute array, the primary mechanism for the establishment of correlations between sites separated by ``short'' distances $\delta r < 1/f$ (i.e., less than the average interparticle spacing) are direct interactions between the Rydberg atoms, as the intervening sites between the pair of atoms are typically empty. For dipolar interactions, this leads to a predicted timescale of $t\propto J^{-1}\delta r^3$ for correlations to emerge, which is consistent with the observed nonlinear lightcone [see green lines in panel (b)]. Similar sub-ballistic propagation of correlations at distances $\delta r \lesssim 1/f^{1/2}$ is observed in $2$D arrays up to $Jf^{3/2} t \lesssim 1$, as illustrated in \fref{correlations}~(c-d) ($f = 0.01$). In these panels the correlations are computed with $\mathbf{\delta r}$ fixed to lie along the $x$ ($\mathbf{\delta r} = \delta x \hat{\mathbf{x}}$) and $y$ ($\mathbf{\delta r} = \delta y \hat{\mathbf{y}}$) axes, respectively. Both panels are for anisotropic interactions ($\phi=\pi/2,\varphi=0$).

\begin{figure}[tb]
		\centering
		\epsfig{file=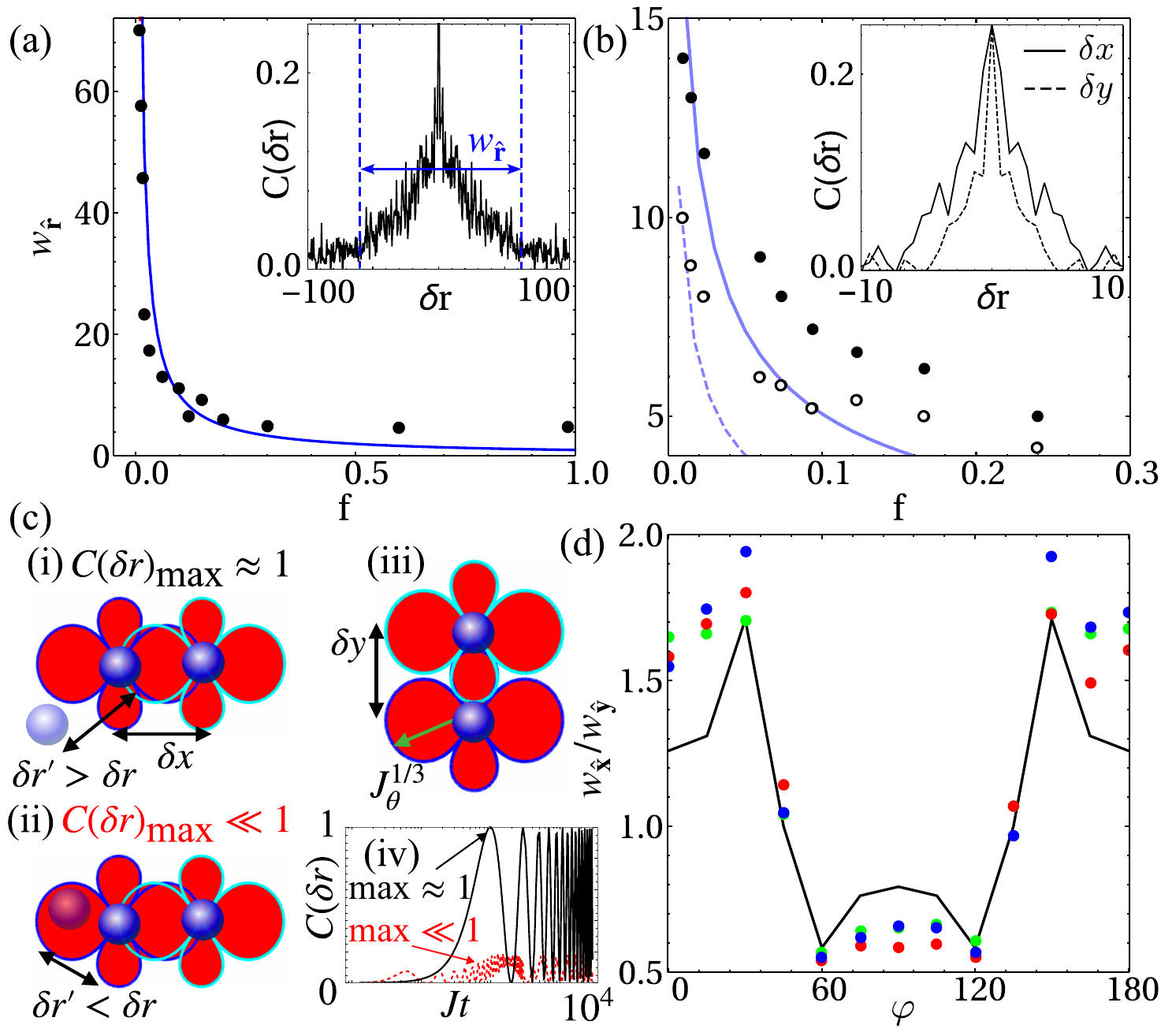,width=1.03 \linewidth}
		\caption{
  (a) Long-time correlation width $w_{\hat{\mathbf{r}}}$ as a function of filling fraction $f$ in (a) $1$D and (b) $2$D arrays. In the latter, we distinguish correlations along the $x$ (solid circles) and $y$ (open circles) axes obtained from numerical simulations. In both panels (a) and (b), lines indicated a fit of the data to the function $N_0/f^{1/d}$ for a constant $N_0$ [solid and dashed lines distinguish correlations along $x$ and $y$, respectively, in (b)]. Insets in both panels illustrate example correlation functions at $Jf^{3/d}t = 10^2$ obtained from data in Fig.~\ref{correlations}~(b-d). 
  (c) Schematic diagram indicating a pair with excluded regions in red dipolar patterns, accompanied by neighboring atoms, where (i) $\delta r'>\delta r$ does not affect the correlations, while (ii) if neighboring atoms are closer $\delta r'<\delta r$ then correlation build up is suppressed between the pairs. (iii) Excluded regions for spins placed along $y$ direction. (iv) Correlation dynamics between spins separated by $\delta r$ for additional spin at $\delta r'$ with $\delta r'>\delta r$ (black) and $\delta r'<\delta r$ (red). (d) Ratio of widths ($w_{\hat{\mathbf{x}}}/w_{\hat{\mathbf{y}}}$) along $x$ and $y$ at different filling fraction $f=0.015$ (green-dots), $f=0.012$ (red-dots) and $f=0.009$ (blue-dots) as a function of quantization axis angle $\varphi$. Solid line indicates analytic prediction (see main text).
			\label{correlations_longtime}}
	\end{figure}

	%
Beyond the initial sub-ballistic regime, we observe that at longer times $J f^{3/d} t \gtrsim 1$ the propagation of correlations is abruptly halted for disordered arrays. Specifically, we find that the pairwise correlations only become appreciably non-zero over distances approximately given by the average inter-particle spacing $1/f^{1/d}$ [see vertical white-dashed lines shown for emphasis in \fref{correlations}~(b) for the $1$D case]. This is to be contrasted with the previously noted spreading of correlations throughout the entire lattice when $f=1$. 
The apparent lack of correlations at longer length scales is systematically verified by extracting the correlation width $w_{\hat{\mathbf{r}}}$, defined as the full-width half maximum (FWHM) of $C(\mathbf{\delta r}, t)$ at long times $Jf^{3/d}t \gg 1$, as a function of filling fraction $f$, as shown in \fref{correlations_longtime}~(a) and (b) (black open and solid markers) for $1$D and $2$D arrays, respectively. Here, $\hat{\mathbf{r}}$ indicates the direction in the $d$-dimensional plane along which we observe the correlation dynamics. 
For low filling fractions ($f \ll 1$) we observe that the correlation width scales as $w_{\hat{\mathbf{r}}} \sim 1/f^{1/d}$ (i.e., much smaller than $L$ in $1$D), whereas for larger filling fractions ($f \sim 1$) the correlation width saturates to a constant set by the system size $L$ ($\sim N$), consistent with expectations for an ordered array. We note that in $2$D the correlation widths at low filling fraction appear to differ by a prefactor, which we explain in the following. 

	%

Unlike the initial build-up of short-range ($\delta r \lesssim 1/f^{1/d}$) correlations, which can be entirely understood via a pair of interacting spins, the suppression of correlations at longer distances arises due to the presence of additional neighboring spins. To see this, consider again the pair-neighbour toy model in $1$D, defined here as a pair of spins separated by distance $\delta r$ accompanied by a third neighbouring spin located at a distance $\delta r'$ away from one of the spins within the original  pair. Depending on the relative distance of the third spin, the dynamics of the pair changes greatly. First, if $\delta r'\gg \delta r$, as depicted in \fref{correlations_longtime}~(c)~(i), the third spin interacts relatively weakly with the pair (i.e., $J/\delta r^3 \gg J/\delta r^{\prime 3}$) and any contributions to the pair dynamics will only arise on timescales much longer than that associated with the build up of correlations, $J t\propto \delta r^3$. On the other hand, if $\delta r' \lesssim \delta r$, as illustrated in \fref{correlations_longtime}~(c)~(ii), the interaction with the third spin becomes the dominant contribution to the dynamics of the pair (i.e., $J/\delta r^3 \ll J/\delta r^{\prime 3}$). In particular, the relative disparity of the energy scales associated with interactions at distances $\delta r'$ and $\delta r$ leads to a strong suppression of the growth of the correlation between the original pair as shown in \fref{correlations_longtime}~(c)~(iv).


The pair-neighbour toy model can be connected to the observed build up or suppression of the correlations at short and long distances, respectively, by considering the probability for a pair of spins to be isolated or feature a strongly interacting neighbouring spin in a particular disorder configuration. To be concrete, in a dilute $1$D lattice, the probability for a pair of spins to be isolated (i.e., not feature a third spin within a distance $\delta r' < \delta r$ of the pair) can be shown to be $P(\delta r)\approx(1-f)^{3\delta r-1}$ (see Appendix~\ref{app:three-spin} for further details). This probability is near unity when $\delta r \ll 1/f$, indicating that correlations at these short distances should be well captured by a model of just two interacting spins, consistent with our prior analysis of \fref{correlations}. On the other hand, the probability essentially vanishes when $\delta r \gtrsim 1/f$, indicating that the contribution from neighbouring spins becomes dominant, which explains the observed suppression of correlations over these longer distances.

 
\section{Impact of anisotropy on correlation propagation}
In $2$D arrays, the suppression of correlations at long distances additionally depends on the anisotropy of the dipolar interaction, which results in correlation widths along $x$- and $y$ that differ by a prefactor. In particular, when extending the interpretation of the previous toy model one must now account for the relative orientation of the pair and neighbouring spins, as it controls the relative strength of the distinct interactions between the pair and neighbouring spin. To see this, consider the pair-neighbor toy model within a $2$D plane such that the quantization axis is set along $\hat{\mathbf{n}}\equiv\hat{\mathbf{x}}$ [$\varphi=0$, see Fig.~\ref{sketch_setup}(b)]. The interatomic vector of the pair is given by $\delta \mathbf{r}$, while an additional spin is positioned away from the pair with separation defined by $\delta \mathbf{r}'$. As discussed earlier, the suppression of correlations depends on the relative interaction strength between the pair ($J_{p}=J(1-3(\delta \mathbf{r}\cdot \hat{\mathbf{n}})^2)/|\delta \mathbf{r}|^3$) and the neighbor with an atom of the pair ($J_{pn}=J(1-3(\delta \mathbf{r}'\cdot \hat{\mathbf{n}})^2)/|\delta \mathbf{r}'|^3$). The region of the array where the presence of the neighboring spin can suppress the correlation build-up ($J_{pn} > J_{p}$) thus now depends on both the distance $\delta r$ and subtly on the relative orientation of the pair (through $\delta\mathbf{r}\cdot \hat{\mathbf{n}}$).
This is illustrated for the cases $\delta \mathbf{r}=\delta x \hat{\mathbf{x}}$ or $\delta \mathbf{r}=\delta y \hat{\mathbf{y}}$, as shown by the representative diagrams (see red regions indicating $J_{pn} > J_{p}$) of (i) and (iii), respectively, in \fref{correlations_longtime}~(c). The probability to occupy this region can be computed in terms of the area $A(\delta r)$ as $P(\delta r)\approx(1-f)^{A(\delta r)}$, and an estimate of $A(\delta r)$ enables us to deduce the scaling of the correlation width as $w_{\hat{\mathbf{r}}}\propto (1-3(\hat{\mathbf{r}}\cdot \hat{\mathbf{n}})^2)^{1/3}/f^{1/2}$ along the axis $\hat{\mathbf{r}}$.
We validate this prediction by computing the ratio of correlation widths $w_{\hat{\mathbf{x}}}/w_{\hat{\mathbf{y}}} =((1-3(\hat{\mathbf{x}}\cdot \hat{\mathbf{n}})^2)/(1-3(\hat{\mathbf{y}}\cdot \hat{\mathbf{n}})^2))^{1/3}=(J_x/J_y)^{1/3}$ in \fref{correlations_longtime}(d) as a function of the azimuthal angle $\varphi$ of the quantization axis in the plane of the array, for a few values of the filling fraction. We observe reasonable qualitative agreement in all cases.

\section{Relaxation of correlations}\label{sec:NN_correlations} While we have shown that the initial build-up of short-range correlations can be well understood through the lens of a simple pair model, the pair-neighbour description can still provide insight into the subsequent relaxation of correlations at long times. In the following we will focus on nearest-neighbour correlations without loss of generality. To avoid fast oscillating contributions to the dynamics of the closely spaced pair, which can obfuscate details of the long-time relaxation behaviour, we consider the lattice-averaged polarization correlation, 
\begin{eqnarray}
		C_{xx}(\delta \mathbf{r},t) &=&\frac{1}{4fL^d}\sum_\mathbf{r} (\langle \hat{\sigma}^{x}_{\mathbf{r}} \hat{\sigma}^{x}_{\mathbf{r}+\delta \mathbf{r}}\rangle - \langle \hat{\sigma}^{x}_{\mathbf{r}} \rangle \langle \hat{\sigma}^{x}_{\mathbf{r}+\delta \mathbf{r}}\rangle).\label{correlation_NN}
\end{eqnarray}

\begin{figure}[tb]
		\centering
		\epsfig{file=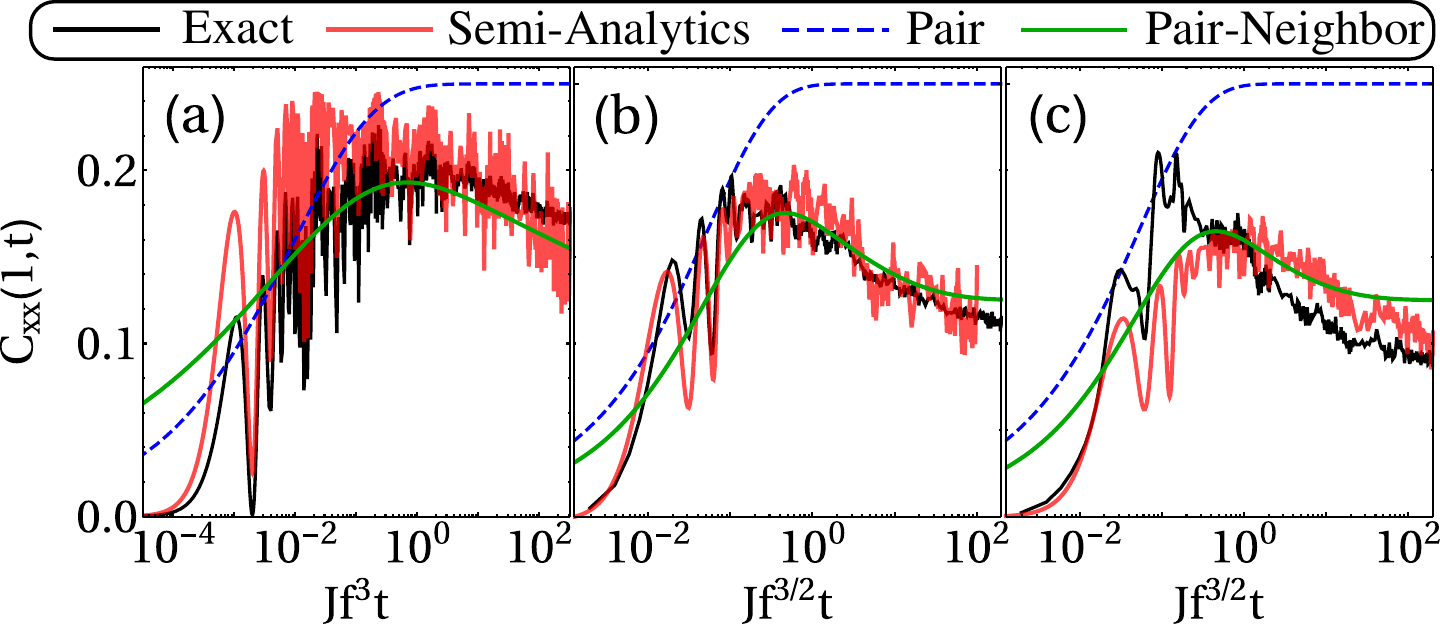,width= \linewidth}
		\caption{Nearest neighbor correlations in (a) 1D ($N=6$) and 2D arrays ($N=9$) along (b) $x$-axis and (c) $y$-axis.  We compare results from full numerical simulations (black lines) as well as analytics from the pair (blue-dashed lines) and pair-neighbour (green lines) models (see text for details). Additionally, we show the numerical solution of the pair-neighbour model without assumptions and using the same disorder realizations as the full numerics (red lines).  All data uses filling fraction $f=0.1$. 
			\label{Cnn}}
\end{figure}

We show results for the nearest-neighbour correlation ($\delta r = 1$) for a $1$D array in \fref{Cnn}~(a), obtained from numerical simulations (black lines) with $f = 0.1$. Initially, the correlation grows from zero to a maximum value at $Jf^3t \sim 1$, before a subsequent period of much slower decay. The first period of growth, which occurs on fast timescales, is dominated by pairs separated by less than the average inter-particle spacing. The previously discussed pair model should thus provide a sufficient description and predicts $\supp{C}{pair}_{xx} =  \left[ 1-\exp(-2(\gamma_p t)^{\beta_p}) \right]/4$ with $\beta_p = 1/3$ as discussed previously for $1$D. Note that the pair model enforces that $\langle \hat{\sigma}^{x}_{r} \hat{\sigma}^{x}_{r+1}\rangle = 1$ remains constant and thus the correlation dynamics are driven by the decay of one-body observables [i.e., the contribution $\langle \hat{\sigma}^{x}_{\mathbf{r}} \rangle \langle \hat{\sigma}^{x}_{\mathbf{r}+1}\rangle$ in Eq.~(\ref{correlation_NN})]. We find excellent agreement between $\supp{C}{pair}_{xx}$ (blue dashed lines) and the numerical results at short times $Jf^3t \lesssim 1$. 



At long times, the pair model predicts the correlation to saturate to a value of $1/4$, which is clearly inconsistent with the numerical results. 
A description of the relaxation at long times thus requires the inclusion of additional spins, which is minimally achieved by using the pair-neighbour model. Analytic insight can be obtained by approximately solving the pair-neighbour model with the assumption that the third neighbouring spin is separated by a large distance from the original pair (see Appendix \ref{app:three-spin}). This yields a tractable expression involving a sum of stretched exponentials,
\begin{eqnarray}
		C^{pn,\mathrm{1D}}_{xx}(\delta \mathbf{r},t) &=&\frac{1}{8}\bigg(1+\exp{[-(\kappa_{pn}^{\mathrm{1D}} |\delta \mathbf{r}|^3 t)^{\frac{1}6}]} \nonumber \\
       & &- \exp{[-2(\gamma_p^{\mathrm{1D}} t)^{\frac{1}3}-2(\gamma_{pn}^{\mathrm{1D}} t)^{\frac{1}6}]} \bigg),
  \label{Cxx_1D}
\end{eqnarray}
where $\kappa^{\mathrm{1D}}_{pn} \propto J f^6 $ is a new two-body relaxation rate that arises from the two-body term in Eq.~(\ref{correlation_NN}) (see \aref{app:three-spin} for details). A comparison between $C^{pn,\mathrm{1D}}_{xx}$ (green line) and the full numerical results in \fref{Cnn}~(a) demonstrates good agreement, including at long times $Jf^3 t \gg 1$. 
For completeness, we also show the results of the numerically evaluated pair-neighbour model (red lines) obtained by averaging over the same disorder configurations as the fully many-body calculations and without the aforementioned assumption about the positioning of the third spin, which qualitatively captures the observed fast oscillations that are absent from our analytic expression.

In 2D arrays [see \fref{Cnn}~(b) and (c)], the dynamics are more complex as they feature multiple relaxation timescales due to the anisotropy of the dipolar interaction. Over short times $Jf^{3/2}t \lesssim 1$, the correlation build-up appears similar along $x$ and $y$ (see black line for many-body calculation), but is slightly faster for the former. This is not captured by a comparison to a `pair' model (blue dashed lines), which in this case corresponds to evaluation of Eq.~\ref{correlation_NN} with the assumption $\langle \hat{\sigma}^{x}_{\mathbf{r}} \hat{\sigma}^{x}_{\mathbf{r}+\delta \mathbf{r}}\rangle = 1$ while $\langle \hat{\sigma}^{x}_{\mathbf{r}} \rangle$ and $\langle \hat{\sigma}^{x}_{\mathbf{r}+\delta\mathbf{r}} \rangle$ are given by the prior pair-neighbour model expression used in Fig.~\ref{relaxation} that does not distinguish $\delta\mathbf{r}$ along $x$ or $y$. 
At longer times, $Jf^{3/2}t \gtrsim 1$, the relaxation of $C_{xx}$ features a marginally slower decay in panel (b) [correlations along $x$-axis] compared to panel (c) [$y$-axis].
We similarly derive an expression for the correlation from the pair-neighbour model (green lines), which in $2$D is written as
%
\begin{eqnarray}
		C^{pn,\mathrm{2D}}_{xx}(\delta \mathbf{r},t) &=&\frac{1}{8}\bigg(1+\exp{\bigg[-\bigg(\frac{\kappa^{\mathrm{2D}}_{pn} |\delta \mathbf{r}|^3}{|1-3\cos^2 \theta|} t\bigg)^{\frac{1}3}}\bigg] \nonumber\\
  & &- \exp{[-2(\gamma_p^{\mathrm{2D}} t)^{\frac{2}{3}}-2(\gamma_{pn}^{\mathrm{2D}} t)^{\frac{1}{3}}}] \bigg),
  \label{Cxx_2D}
\end{eqnarray}
with two-body relaxation rate $\kappa^{\mathrm{2D}}_{pn}\propto Jf^3$. The relative axis of the correlation now appears in the first stretched exponential (via the dependence on $\theta$) and does predict a slightly slower decay at long times, which qualitatively explains observations of the full many-body calculations. 

	%

More generally, for $d$-dimensional system with power-law interactions we find the two-body relaxation rate $\kappa_{pn}\propto Jf^{d/2\alpha}$ and associated stretch exponents $\beta_p=d/\alpha$ and $\beta_{pn}=d/2\alpha$.  We highlight that these stretch exponents (including those in \eref{Cxx_1D} and \eref{Cxx_2D}) are specific to the XY Hamiltonian, which demonstrates that the glassy relaxation of correlations, even in $1$D, can be a strongly distinguishing feature between systems described by different spin models. Lastly, we also remark that the pair-neighbour model predicts a distinct maximum in the correlation at $Jf^{3/d}t \sim 1$ followed an eventual saturation to $C^{pn}_{xx} \to 1/8$, which strikingly distinguishes it from the pair model which grows toward a long time value of $C^{\mathrm{pair}}_{xx} \to 1/4$. This contrasts with one-body observables such as the collective polarization $\langle \hat{S}_x \rangle$, which is predicted to identically vanish at long times for both pair and pair-neighbour models, and demonstrates that correlation functions could be a more sensitive probe of the role of three or more spins in the relaxation dynamics even in $1$D.

 
	\section{Conclusions and outlook}
	%
In this work, we have investigated the relaxation dynamics of arrays of interacting dipoles featuring positional disorder and anisotropic interactions and presented a number of key findings. For example, we have demonstrated that the dynamical phase diagram for $2$D arrays is uniquely enriched by the tunable anisotropy of the dipolar interactions and conflicts with naive expectations. Furthermore, we have shown that two-body correlations can exhibit signatures of the positional disorder, including sub-ballistic growth of short-range correlations at short times, followed by glassy relaxation and the suppression of appreciable long-range correlations at long times. To explain these results we have developed a pair-neighbour model of three interacting spins, which goes beyond prior theoretical treatments based on a simpler pair models. 

Tweezer arrays of dipolar interacting Rydberg atoms present an ideal platform to explore our predictions. In particular, they feature deterministic preparation and can enable systematic scans over the full range of filling fractions, which provide significant advances over prior studies of disordered spin models based on polar molecules in optical lattices \cite{ospelkaus2006ultracold,ni2018dipolar,yan2013observation,hazzard2014many}. Moreover, the deterministic control of positional disorder -- a distinguishing feature from polar molecules in lattices or frozen Rydberg gases \cite{schultzen2022glassy,signoles2021glassy,franz2022observation} -- provides the opportunity to verify details of, e.g., the pair-neighbour model in-situ. In general, the versatility and unique characteristics of tweezer arrays offer a plethora of exciting possibilities for future research in new directions. For instance, high-fidelity site-resolved state preparation (combined with deterministic control) can enable the exploration of alternative initial states, including those featuring arbitrary imprinted spin-order \cite{Covey_2018,dominguez2023relaxation,babadi2015far} or entanglement and correlations \cite{wilk2010entanglement,bluvstein2022quantum,evered2023high}, which can be used to probe beyond the regime of few-spin effective descriptions such as the pair and pair-neighbour models. We note that despite our attention towards Rydberg atoms, rapid technical advances in trapping and manipulating polar molecules in optical tweezers suggest they can also be an exciting platform with similar capabilities \cite{doi:10.1126/science.adf8999,ruttley2023formation,vilas2024optical}. Lastly, we emphasize that the insight provided by our approach can provide an enhanced understanding of other platforms featuring dipolar interactions, including NV centers \cite{kucsko2018critical,davis2023probing,zu2021emergent,hughes2024strongly}.

	\acknowledgments
        We acknowledge stimulating discussions with Arghavan Safavi-Naini and Yicheng Zhang. This material is based upon work supported by the Air Force Office of Scientific Research under award numbers FA9550-22-1-0335.

\appendix
\section{Pair description of XY model \label{app:pair}}
 
	
	%
	\begin{figure}[htb]
		\centering
		\includegraphics[width=0.75\linewidth]{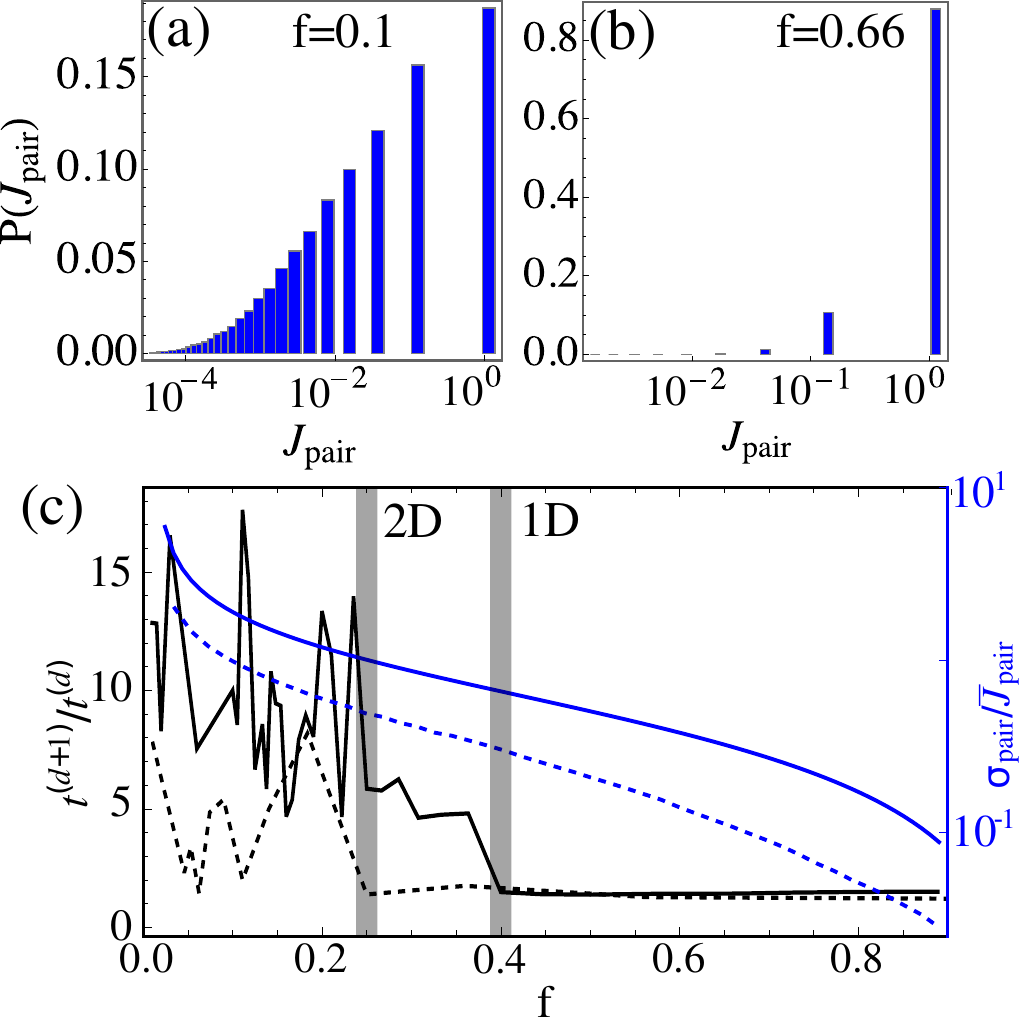}
		\caption{ Distribution statistics of $\sub{J}{pair}$ for (a) $f=0.1$ and (b) $f=0.66$. (c) $t^{(d+1)}/t^{(d)}$ (black) and  $\sub{\sigma}{pair}/\sub{\bar{J}}{pair}$ (blue) with filling fraction $f$, calculated from relaxation dynamics and distribution statistics of 1D (solid) and 2D (dashed) array, respectively, see text.}
		\label{dynamical_phase_diagram_pair}
	\end{figure}

\subsection{Utility of pair model to diagnose dynamical phases}
Analytic insight into glassy dynamics was provided in Ref.~\cite{schultzen2022glassy}, which focused on a model featuring power-law Ising interactions and positional disorder. In that case, the dynamics can be solved exactly \cite{FossFeig2013ising} and in the limit of low filling fraction effectively decomposes into pairs of interacting spins. In particular, the relaxation of the collective polarization can be written as $\langle \hat{S}_x \rangle_{\mathrm{pair}} = \frac{N}{2}\overline{\cos(\sub{J}{pair}t)}$ where $J_{\mathrm{pair}}$ describes the interaction strength of a given pair and the overline $\overline{\cdot \cdot \cdot}$ denotes averaging over all possible pair configurations (i.e., separation distance and orientation) with uniform probability. The averaging for this pair model can be carried out analytically to yield $\langle \hat{S}_x \rangle_{\mathrm{pair}} = \frac{N}{2}\exp[-(\gamma_p t)^{\beta_p}]$ where the stretch exponent $\beta$ for the Ising model is derived as $\beta_p=d/\alpha$.
We note that, as discussed in Ref.~\cite{franz2022observation}, the pair model can be generalized to arbitrary XYZ-type spin interactions and the form of the polarization for each pair is functionally identical. In particular, after averaging over the positional disorder all the models identically predict a stretch exponent $d/\alpha$, as this is dictated by the range of the interaction, but on the other hand one obtains a relaxation rate that scales with $f^{\alpha/d}$) with a model-dependent prefactor.

Despite our observation that the pair model is not always quantitatively applicable for describing the glassy dynamics of the dipolar XY model, it can still provide qualitative evidence to support some of our findings. For example, the statistical distribution of $\sub{J}{pair}$ can be loosely used to predict the parameter regimes in which the different dynamical phases of the dipolar XY model exist.
In \fref{dynamical_phase_diagram_pair}~(a) and (b), we show the distribution of $\sub{J}{pair}$ in a large $1$D system with dipolar interactions for filling fraction $f=0.1$ and $f=0.66$, respectively. In the latter case, the distribution of $\sub{J}{pair}$ is strongly peaked at the nearest-neighbour interaction strength, due to the relatively large filling fraction, which indicates a quite ordered system wherein, e.g., the prediction $\langle \hat{S}_x \rangle_{\mathrm{pair}}$ is clearly inappropriate. On the other hand, when the filling fraction is low, such as $f=0.1$ in panel (a), we observe a large spread in the distribution of $\sub{J}{pair}$ as a result of the strong positional disorder, which implies that $\langle \hat{S}_x \rangle_{\mathrm{pair}}$ can be a reasonable first approximation to the dynamics. As $\langle \hat{S}_x \rangle_{\mathrm{pair}}$ predicts glassy decay of the polarization, we thus conclude that the spread of $\sub{J}{pair}$ values can be used as a loose proxy for the presence or absence of glassy dynamics. 

In \fref{dynamical_phase_diagram_pair}~(c) we show the ratio $\sub{\sigma}{pair}/\sub{\bar{J}}{pair}$ as a function of filling fraction $f$ in $1$D (blue solid line) and $2$D (blue dashed line), where $\sub{\sigma}{pair}$ and $\sub{\bar{J}}{pair}$ are the variance and mean of the distribution of pair interaction strengths. As per the above discussion, the ratio $\sub{\sigma}{pair}/\sub{\bar{J}}{pair}$ provides a metric for the degree of disorder in an array. When $\sub{\sigma}{pair}/\sub{\bar{J}}{pair} \ll 1$ we predict ordered dynamics (in the extreme case of $f=1$ the ratio vanishes), while for $\sub{\sigma}{pair}/\sub{\bar{J}}{pair} \gg 1$ we predict glassy decay. We observe that this metric roughly correlates with the predictions of the main text using the alternative ratio $t^{(d+1)}/t^{(d)}$ (black lines), particularly qualitative features such as the shifting of the crossover between glassy and ordered dynamics to lower values of $f$ as the dimensionality of the array is increased.

\subsection{Breakdown of the pair model\label{app:break_pair}}
As pointed out in the main text, despite the success of the pair model in describing dynamics for the Ising Hamiltonian, it is not sufficient for dipolar XY systems. Here, we systematically explore its applicability as a function of interaction range and array dimension. 

\begin{figure}[htb]
		\centering
		\includegraphics[width=0.85\linewidth]{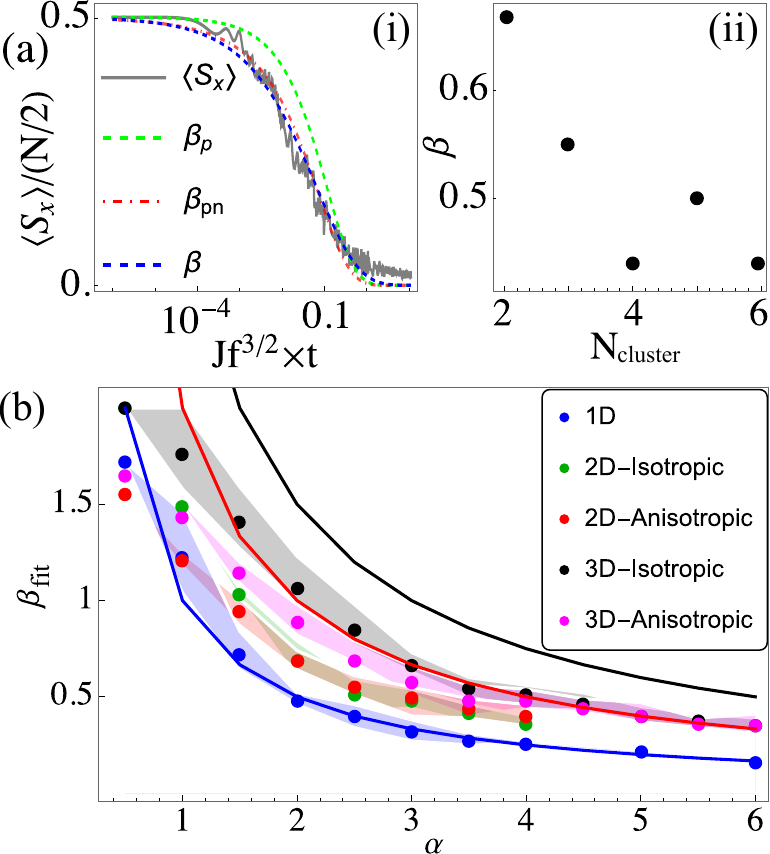}
		\caption{(a)-(i) Time-evolution of ensemble average polarization in 2D array with anisotropic interaction for filling fractions $f=0.01$, compared with pair model (green-dashed), pair-neighbor model (red-dotdashed) and fitted with stretched exponential (blue-dashed). (ii) Variation of fitted stretch exponent $\beta$ with $\sub{N}{cluster}$, where $\sub{N}{cluster}$ indicates a cluster of spins considered in numerical simulations. (b) Variation of fitted stretch exponent $\beta$ with power-law $\alpha$ for 1D array (blue), 2D array with isotropic (green) and anisotropic (red) dipolar interactions, and 3D array with isotropic (black) and anisotropic (magenta) dipolar interactions. Blue-solid line represents $1/\alpha$, while red-solid and black-solid represents $2/\alpha$ and $3/\alpha$, respectively.}
		\label{beta_alpha}
	\end{figure}

In panel (i) of \fref{beta_alpha}~(a), we show example relaxation dynamics of the collective polarization $\langle \hat{S}_x \rangle$ in a 2D array obtained from full many-body calculations with filling fraction $f=0.01$ (gray line). The results are compared with the analytic prediction of the pair model (green dashed), which fails to properly capture the dynamics. This is emphasized by fitting $\langle \hat{S}_x \rangle$ with a generic stretched exponential $(N/2)\exp[-(\gamma t)^\beta]$, where $\gamma$ and $\beta$ are allowed to be free parameters. We obtain a fitted stretch exponent $\beta \approx 0.43$, which is a substantial deviation from the predicted value of ${\beta}_p=2/3)$.




Given the known specificity of the existence of the synchronized phase to $2$D arrays with isotropic dipolar interactions (in general, synchronization with arbitrary power-law interactions requires $\alpha/2 < d < \alpha$ for $d$-dimensional arrays), it is natural to investigate if the breakdown of the pair model that we observe is similarly fine-tuned. To do this, we compute the many-body dynamics of 1D, 2D and 3D systems, with both isotropic and anisotropic interactions, as a function of interaction range $\alpha\in\{0,6\}$. In \fref{beta_alpha}~(b), we present the fitted stretch exponent $\beta$ (markers, see legend) and compare to $d/\alpha$ (solid lines) predicted from the pair description. While we observe good agreement with the pair model in $1$D ($\beta\approx \beta_p$, blue markers and line), we find the pair model fails for $d>1$ across all values of $\alpha$ considered. We note that our results are consistent with prior work based on semiclassical calculations \cite{schultzen2022semiclassical}, which primarily focused on $\alpha = 6$ in $3$D.  Interestingly, for $\alpha \lesssim 3$, we find distinct $\beta$ for isotropic and anisotropic interactions \cite{schultzen2022glassy}, which further emphasizes the important role of interaction anisotropy in the relaxation dynamics of the system.

%

\section{Pair-neighbour description of the XY model\label{app:three-spin}}
		\begin{figure}[htb]
			\centering
			\includegraphics[width=0.99\linewidth]{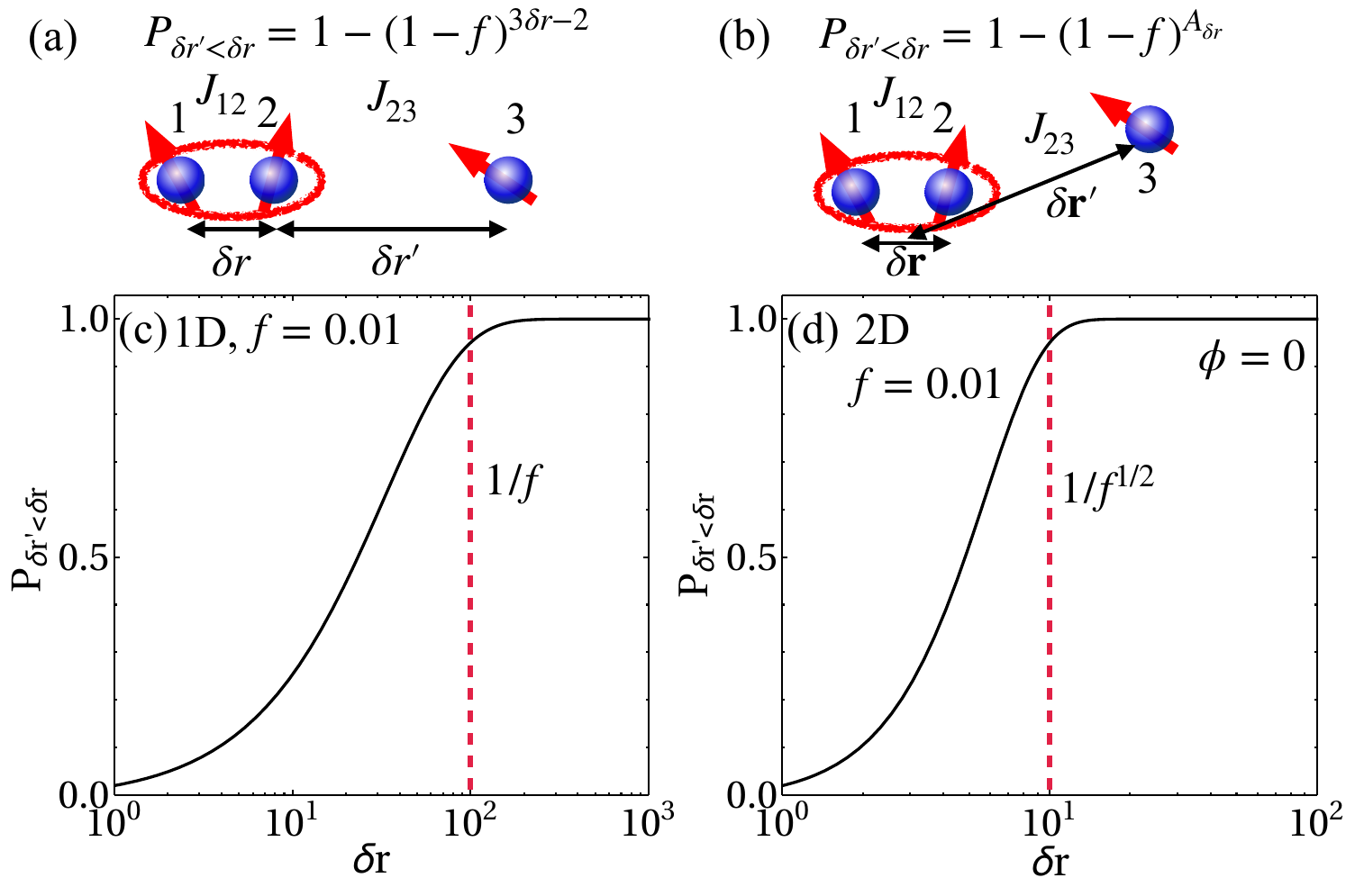}
			\caption{Schematic of a three-spin model in (a) 1D and (b) 2D systems, where spins 1 and 2 for the pair and are separated by distance $\delta r$ and spin 3 is located at a distance $\delta r'$ from the pair. Probability of third spin to be near the pair with distance $\delta r'<\delta r$ in (c) 1D and (d) 2D.}
			\label{P_drp_dr}
		\end{figure}
		
In the previous section and main text, we have highlighted that a pair model is not sufficient to describe the relaxation dynamics of the XY spin Hamiltonian featuring positional disorder. As previously stated, we attribute this to the fact that, unlike the Ising Hamiltonian that has been analytically studied \cite{schultzen2022glassy}, the pairwise interactions in the many-body Hamiltonian do not commute. Thus, in an effort to nevertheless obtain analytic insight into the dynamics of the XY Hamiltonian, we develop a ``pair-neighbour'' model involving three spins. 

We consider a system of three spins [see \fref{P_drp_dr}~(a-b)] composed of a pair (labelled 1 and 2) and a neighbour (3). The pair is characterized by the separation $\delta \mathbf{r}$ and the neighbouring spin is separated by $\delta \mathbf{r}'$ from one of the spins in the pair. To enable the attainment of an analytic expression, we ignore the anisotropy of the dipolar interaction and assume that 
each spin in the pair interacts with the neighbour spin with the same strength, i.e., $J_{23}\approx J_{13}$ and $J_{12}\gg J_{23}$, where we abuse prior notation and denote $J_{ij}$ as the interaction between spins $i$ and $j$ in the triplet. Though this approximation is only strictly valid when $\delta r' \gg \delta r$, it enables us to capture first corrections beyond the pair model. 
For a single configuration of the spins, the time-evolution of the local spin polarization is,
		\begin{eqnarray}
			\langle \hat{\sigma}^x_{1} (t) \rangle &=& \langle \hat{\sigma}^x_{2} (t) \rangle, \nonumber\\
			&=& \bra{ \psi_0} e^{i\sub{\hat{H}}{XY}t} \hat{S}_x^{(1)} e^{-i\sub{\hat{H}}{XY}t} \ket{\psi_0},\nonumber\\
			&\approx&\frac{1}{2}\bigg[\cos\bigg(J_{12}t+\frac{2J_{23}^2}{J_{12}}t\bigg)\nonumber\\ 
   & &+\cos\bigg(J_{12}t+\frac{4J_{23}^2}{J_{12}}t\bigg)\bigg], \\
			 \langle \hat{\sigma}^x_{3} (t) &\approx&\frac{1}{2}\bigg[1+\cos\bigg(\frac{2J_{23}^2}{J_{12}}t\bigg) \nonumber \\
    & &+\frac{2J_{23}}{J{12}}\cos\bigg(J_{12}t+\frac{4J_{23}^2}{J_{12}}t\bigg)\bigg] , \label{Sx^3_XY}
		\end{eqnarray}

We obtain the disorder-averaged behavior of the local polarization $\overline{\langle \hat{\sigma}^x_1 (t) \rangle}$ by averaging over all possible configurations of the spins. Switching notation $\delta r \rightarrow r$ and $\delta r'\rightarrow r'$, the average can be obtained in the limit of vanishing filling fraction as
\begin{eqnarray}
		\overline{\langle \hat{\sigma}^x_1 (t) \rangle}&=&\frac{1}{2}\int_V dr dr' ~ P( r) P( r') \bigg[ \cos \bigg(\frac{J t}{ r^\alpha}+\frac{2Jr^{\alpha} t}{ r'^{2\alpha}}\bigg) \nonumber\\ 
  & & + \cos\bigg(\frac{J t}{ r^\alpha}+\frac{4Jr^{\alpha} t}{ r'^{2\alpha}}\bigg) \bigg], \label{S_x_integral}
\end{eqnarray}
where for completeness we have assumed power-law interactions of the form $J_{12} = J/r^\alpha$ and $J_{23} = J/r^{\prime \alpha}$ with exponent $\alpha$, and we have introduced the radial probability $P(r)$ of the second pair (or neighbour) spin to lie within a distance $r$ of the ``central'' spin $1$. To be general, we consider the integration volume $V$ to be a $d$-dimensional sphere of radius $L$. 

In the limit of $f\to0$, the probability of an individual site to be occupied with an atom is $f$ and the probability of an empty site is $(1-f)$. This enables us to simply write down $P(r)$ as the joint probability of two occupied sites separated by a distance $r$ with all intervening sites empty, i.e., $P(r)=2f(1-f)^{2r-1}$ in $1$D and $P(r) = 2\pi r f(1-f)^{\pi(r-1)^2}$ in $2$D. To obtain insightful expressions from Eq.~(\ref{S_x_integral}), we will assuming a vanishingly small filling fraction, $f\to 0$, and simultaneously $L \to \infty$ (such that $f L^d = \mathrm{constant}$). We thus replace the probabilities in Eq.~(\ref{S_x_integral}) with the approximations $P(r)\approx2f$ and $2\pi r f$ in $1$D and $2$D, respectively. Further, within this approximation we re-normalize the probabilities [equivalently Eq.~(\ref{S_x_integral})] with factors of $2fL$ in $1$D and $\pi f L^2$ in $2$D.

 

Carrying out the integral over $r'$ in \eref{S_x_integral} and taking the limit $L\to \infty$ yields
\begin{eqnarray}
		\overline{\langle \hat{\sigma}^x_1 (t) \rangle}&=&\int_{V_p} dr ~ P( r) \bigg\{\cos\bigg(\frac{J t}{ r^\alpha}\bigg)  \nonumber\\ 
  & &+(Jr^{\alpha} t)^{d/2\alpha}f^d\Gamma\bigg(-\frac{d}{2\alpha}\bigg)\bigg[{\cal C}_1^{d,\alpha} \cos\bigg(\frac{J t}{r^\alpha}\bigg)\nonumber\\ 
  & &+{\cal C}_2^{d,\alpha} \sin\bigg(\frac{J t}{r^\alpha} \bigg) \bigg]\bigg\},
\end{eqnarray}
where $V_p$ is the remaining integration volume associated with the pair and we have introduced the constants ${\cal C}_1^{d,\alpha}$ and ${\cal C}_2^{d,\alpha}$, which in the relevant cases of $\alpha = 3$ and $d = 1,2$ are given by:
\begin{align*}
    \mathcal{C}_1^{1,3}&=\frac{\sqrt{3}+1}{2^{7/6}}, &\mathcal{C}_2^{1,3}&=\frac{\sqrt{3}-1}{2^{7/6}}, \\
    \mathcal{C}_1^{2,3}&=2^{4/3}\sqrt{3},&\mathcal{C}_2^{2,3}&=2^{4/3} .
\end{align*}
  
Solving the remaining integral over $r$ and taking the limits $L\to \infty$ and $f\to 0$, we obtain
  \begin{eqnarray}
  \overline{\langle \hat{\sigma}_1^x (t) \rangle}& \approx 1 - (\gamma_p^{d,\alpha} t)^{d/\alpha} - (\gamma_{pn}^{d,\alpha} t)^{d/2\alpha}, \nonumber \\ 
  &\approx  \exp{[- (\gamma_p^{d,\alpha}t)^{\frac{d}{\alpha}}- (\gamma_{pn}^{d,\alpha} t)^{\frac{d}{2\alpha}}}],
  \label{S_x_integral_r}
\end{eqnarray}
with relaxation rates,
\begin{eqnarray}
    \gamma_p^{d,\alpha}&=&\bigg[-\frac{J^{d/\alpha}f}{2^{2-d}\alpha^{d/2}}\Gamma\bigg(-\frac{d}{\alpha}\bigg)  \bigg]^{\alpha/d}, \label{eqn:gamma_p} \\
    \gamma_{pn}^{d,\alpha} &=&\bigg[-\frac{\mathcal{C}_1^{d,\alpha}J^{\frac{d}{2\alpha}}\sqrt{f}\pi^{\frac{d-1}2}}{\alpha 2^{d/2}}\Gamma\bigg(-\frac{d}{2\alpha}\bigg)\bigg]^{2\alpha/d}, \label{eqn:gamma_pn}
\end{eqnarray}
derived specifically for the cases of $\alpha = 3$ and $d = 1,2$. We note that the pair-neighbour model and its predictions for the stretch exponents are, distinct from the pair model, specific to the XY interaction.

We emphasize that because our expression \eref{S_x_integral_r} is obtained under the assumption $\delta r'\gg \delta r$, which we abuse during the solution of the integral (\ref{S_x_integral}), it should only be treated as an approximation that provides qualitative insight into the results of our many-body simulations. 
In the main text, the disorder-averaged collective polarization is plotted from the final result (\ref{S_x_integral_r}) as $\overline{\langle \hat{S}_x(t) \rangle} = (N/2)\overline{\langle \hat{\sigma}^x_1(t) \rangle}$ with relaxation rates $\gamma_p^{d,\alpha}$ and $\gamma_{pn}^{d,\alpha}$ that differ in the case of $2$D from the expressions (\ref{eqn:gamma_p}) and (\ref{eqn:gamma_pn}) by a prefactor of $\pi/2$ and are obtained by fitting to the many-body calculation results. This rescaling will additionally have a contribution from the anisotropy of the dipolar interactions that we have neglected in our derivation \cite{schultzen2022glassy}.

In fact, improved quantitative predictions (particularly at long times) require not only the relaxation of assumptions on the spin-spin couplings, but also additional spins, both of which quickly make the problem intractable.
For illustration, in \fref{beta_alpha}~(a)-(ii) we compare results of numerical simulation with $\sub{N}{cluster} \leq N$ spins in a 2D array with filling fraction $f=0.01$ [i.e., the same conditions as panel (i) where $N = 9$]. It can be seen that the fitted stretch exponent $\beta$ varies across the range $\sub{N}{cluster}\in\{2,6\}$, but tends towards $0.44$ for $\sub{N}{cluster}>3$, which is consistent with the result obtained in panel (i) with $N = 9$. 

Beyond the decay of the collective polarization, the pair-neighbour model can also be used to obtain expressions for the polarization correlation $C_{xx}(\delta\mathbf{r},t)$ used in Fig.~\ref{Cnn} of the main text. 
The two-body term in $C_{xx}(\delta\mathbf{r},t)$ is obtained for a single disorder configuration as,
\begin{eqnarray}
	\langle \hat{\sigma}^x_1 \hat{\sigma}^x_2\rangle &=&  \bra{ \psi_0} e^{i\sub{\hat{H}}{XY}t} \hat{\sigma}_x^{(1)}|_r \hat{\sigma}_x^{(2)}|_{r+\delta r} e^{-i\sub{\hat{H}}{XY}t} \ket{\psi_0},\nonumber\\
	&\approx&\frac{1}{8}\bigg[1+\frac{3J_{23}}{2J_{12}}\bigg(1+\cos(J_{12}t)\bigg) \nonumber\\
	& &+ \bigg(1-\frac{2J_{23}}{J_{12}}\bigg)\cos\bigg(\frac{2J_{23}^2}{J_{12}}t\bigg) \nonumber \\
	& &-\frac{2J_{23}}{J_{12}}\cos\bigg(J_{12}t+\frac{2J_{23}^2}{J_{12}}t\bigg) \bigg], \\
	&=&\frac{1}{2}\bigg(1+\cos\bigg(\frac{2J_{23}^2}{J_{12}}\bigg)\bigg) ,
	\end{eqnarray}
where we again assumed that $J_{12}\gg J_{23} = J_{13}$. We average over the location of the neighbour spin in the limit of vanishing filling fraction by computing the integral, 
\begin{equation}
    \overline{\langle \hat{\sigma}^x_1 \hat{\sigma}^x_2  \rangle} = \int_V dr dr' P(r) P(r') ~ \langle \hat{\sigma}^x_1 \hat{\sigma}^x_2 \rangle , 
\end{equation}
with $J_{12} = J(1-3(\delta \mathbf{r}\cdot \hat{\mathbf{n}})^2)/|\delta \mathbf{r}|^3$ and $J_{23} = J/\delta r'^3$. 
Following similar steps as per \eref{S_x_integral}, we obtain
\begin{flalign}
	\overline{\langle \hat{\sigma}^x_1 \hat{\sigma}^x_2  \rangle} &= \frac{1}{2}+\frac{1}{2}\exp{\bigg[-\bigg(\frac{\kappa_{pn}^{d,\alpha}|\delta \mathbf{r}|^3 t}{|1-3(\delta \mathbf{r}\cdot \hat{\mathbf{n}})^2|}\bigg)^{\frac{d}{2\alpha}}\bigg]},
\end{flalign}
where $\kappa_{pn}^{d,\alpha}=\gamma_p^{d,2\alpha}/2^{d/\alpha}$.

\section{Correlation widths at long times\label{app:correlation_pn} } 
In the main text, we associated the initial propagation of short-range correlations with the dynamics of isolated pairs of spins, while at long times we argued that the suppression of correlations at large distances was a result of sufficiently strong interactions with a third neighbouring spin. A crucial part of this discussion is determining the probability for a disorder realization featuring the latter configuration. 

A schematic of the relevant pair-neighbour configuration in $1$D is illustrated in \fref{P_drp_dr}~(a). As per the discussion of the prior sections, we consider a pair of spins (labeled $1$ and $2$), with separation $\delta r$, accompanied by an additional spin (labeled $3$) separated by $\delta r'$ from one member of the pair. Without loss of generality we assume that spin $3$ is located closer to spin $2$ such that in $1$D the associated pair and pair-neighbour interactions are, following the notation introduced in the previous section, $J_{12}=J/\delta r^3$ and $J_{23} = J/\delta r'^3$, respectively. 
We seek to determine the probability $P$ that the neighbouring spin is more strongly interacting than the pair, $J_{23} > J_{12}$, or equivalently that the neighbouring spin is at a distance $\delta r' \leq \delta r$. This is most easily obtained by computing the probability of the complementary event, which corresponds to all sites within a range of $\delta r$ of the pair being unoccupied, i.e., i) $\delta r$ sites to the left of the pair are unoccupied, ii) $\delta r$ sites to the right of the pair are unoccupied and iii) the $\delta r - $ sites between the pair are unoccupied. For vanishing filling fraction $f \to 0$ the probability for a single site to be unoccupied is $(1-f)$, and thus we have $P(\delta r) = 1 - (1-f)^{3\delta r - 1}$.


A similar computation can be carried out for $2$D [illustrated in \fref{P_drp_dr}~(a)], but now the anisotropy of the respective interactions must be accounted for. In particular, determining the probability $P$ that $J_{pn} > J_p$ depends on determining not just the relative distance of the neighbour (i.e., comparing $\delta r$ and $\delta r'$, but also the relative orientations of the pair and neighbour (i.e., $\theta$ and $\theta'$). It is straightforward to determine that this probability can be expressed in the general form $P(\delta r, \theta) = 1 - (1-f)^{A(\delta r, \theta)}$ where $A(\delta r, \theta)$ is the area surrounding the pair such that $J_{23} > J_{12}$ [illustrated in \fref{correlations_longtime}(c) of the main text]. An analytic expression for $A(\delta r, \theta)$ is difficult to obtain, but on general grounds one can argue that it should approximately behave as $A(\delta r, \theta) \propto \delta r^2 [1-3(\delta \mathbf{r}\cdot \hat{\mathbf{n}})^2]^{-2/3}$ and we have confirmed this with numerical calculations. 

In Figs.~\ref{P_drp_dr}(c) and (d) we plot examples of the respective probabilities for $1$D and $2$D as a function of $\delta r$, with the choice of $\phi=0$ in the latter so that the interactions are isotropic. We observe that the probability of having at least one spin near a pair quickly saturates to unity when $\delta r $ exceeds the average interparticle spacing $1/f$ in $1$D and $1/f^{1/2}$ in $2$D, and thus following the discussion of the main text we expect correlations between pairs separated by these distances to be suppressed. In particular, this leads us to predict the correlation widths to scale as $w_{\hat{\mathbf{r}}} = f$ and $w_{\hat{\mathbf{r}}} = f^{1/2}[1-3(\hat{ \mathbf{r}}\cdot \hat{\mathbf{n}})^2]^{1/3}$ in $1$D and $2$D, respectively.

 \newpage
	
	%

	
\end{document}